\documentclass[british,american,aps,prb,reprint,superscriptaddress,preprintnumbers,showpacs,sectcntreset]{revtex4-1}
\usepackage{lmodern}

\usepackage[T1]{fontenc}
\usepackage[latin9]{inputenc}
\usepackage{xcolor}
\usepackage{pdfcolmk}
\usepackage{babel}
\usepackage{float}
\usepackage{textcomp}
\usepackage{amsmath}
\usepackage{amssymb}
\usepackage{graphicx}
\PassOptionsToPackage{normalem}{ulem}
\usepackage{ulem}
\usepackage[unicode=true,pdfusetitle,
 bookmarks=true,bookmarksnumbered=false,bookmarksopen=false,
 breaklinks=false,pdfborder={0 0 1},backref=false,colorlinks=false]
 {hyperref}

\makeatletter

\providecommand{\tabularnewline}{\\}
\providecolor{lyxadded}{rgb}{0,0,1}
\providecolor{lyxdeleted}{rgb}{1,0,0}

\usepackage{upgreek}

\makeatother

\begin{document}

\title{High-fidelity initialization of long-lived quantum dot hole spin
qubits by reduced fine-structure splitting}

\author{A. J. \surname{Brash}}

\author{L. M. P. P. \surname{Martins}}

\author{F. \surname{Liu}}

\email[To whom correspondence should be addressed: ]{ FengLiu@sheffield.ac.uk}

\selectlanguage{american}%

\affiliation{Department of Physics and Astronomy, University of Sheffield, Sheffield,
S3 7RH, United Kingdom}

\author{J. H. \surname{Quilter}}

\affiliation{Department of Physics and Astronomy, University of Sheffield, Sheffield,
S3 7RH, United Kingdom}

\affiliation{Department of Physics, Royal Holloway, University of London, Egham,
TW20 0EX, United Kingdom}

\author{A. J. \surname{Ramsay}}

\affiliation{Hitachi Cambridge Laboratory, Hitachi Europe Ltd., Cambridge CB3
0HE, United Kingdom}

\author{M. S. \surname{Skolnick}}

\author{A. M. \surname{Fox}}

\affiliation{Department of Physics and Astronomy, University of Sheffield, Sheffield,
S3 7RH, United Kingdom}

\date{\today}

\begin{abstract}
We demonstrate an on-demand hole spin qubit initialization scheme
meeting four key requirements of quantum information processing: fast
initialization (1/e \textasciitilde{} 100 ps), high fidelity ($F>99\%$),
long qubit lifetime ($2T_{h}>T_{2}^{*}\simeq10\:\mathrm{ns}$), and
compatibility with optical coherent control schemes. This is achieved
by rapidly ionizing an exciton in an InGaAs quantum dot with very
low fine-structure splitting at zero magnetic field. Furthermore,
we show that the hole spin fidelity of an arbitrary quantum dot can
be increased by optical Stark effect tuning of the fine-structure
splitting close to zero.
\end{abstract}
\maketitle
Single hole spins confined in semiconductor quantum dots (QDs) are
an attractive stationary qubit candidate owing to their long coherence
times \citep{Brunner03072009,DeGreve2011,PhysRevLett.108.017402},
ultrafast optical coherent control \citep{DeGreve2011,Greilich2011,PhysRevLett.108.017402}
and potential for integration with circuit-style devices for quantum
information processing (QIP) \citep{Gao2012,DeGreve2012,:/content/aip/journal/apl/104/23/10.1063/1.4883374}.
Initialization of a qubit to a well-defined state is a critical part
of any QIP protocol as it limits the fidelity of the entire process.
An ideal initialization scheme should be fast, operate on-demand and
have high fidelities to permit error correction \citep{DiVincenzo2000,Preskill1998},
whilst long qubit lifetimes are desirable to maximize the number of
possible gate operations.

A range of single carrier spin initialization schemes have previously
been demonstrated for both single QDs and quantum dot molecules. These
include optical pumping \citep{Atature28042006,PhysRevLett.99.097401,PhysRevLett.101.236804},
coherent population trapping \citep{Brunner03072009,Xu2008} and the
ionization of an exciton \citep{Kroutvar2004,1367-2630-9-10-365,PhysRevLett.100.197401,PhysRevB.77.235442,PhysRevB.85.241306}.
Optical pumping methods have reached fidelities as high as 99.8\%
in an out-of plane magnetic field \citep{Atature28042006} with initialization
times of the order of $\upmu$s. Faster (ns) initialization with slightly
lower fidelities has been observed in an in-plane magnetic field \citep{PhysRevLett.99.097401,Xu2008}.
However, practical fault-tolerant QIP implementations \citep{DiVincenzo2000,Preskill1998}
require initialization that is very fast compared to decoherence and
hence it is desirable to further increase the initialization speed.

\begin{figure}
\includegraphics[width=1\columnwidth]{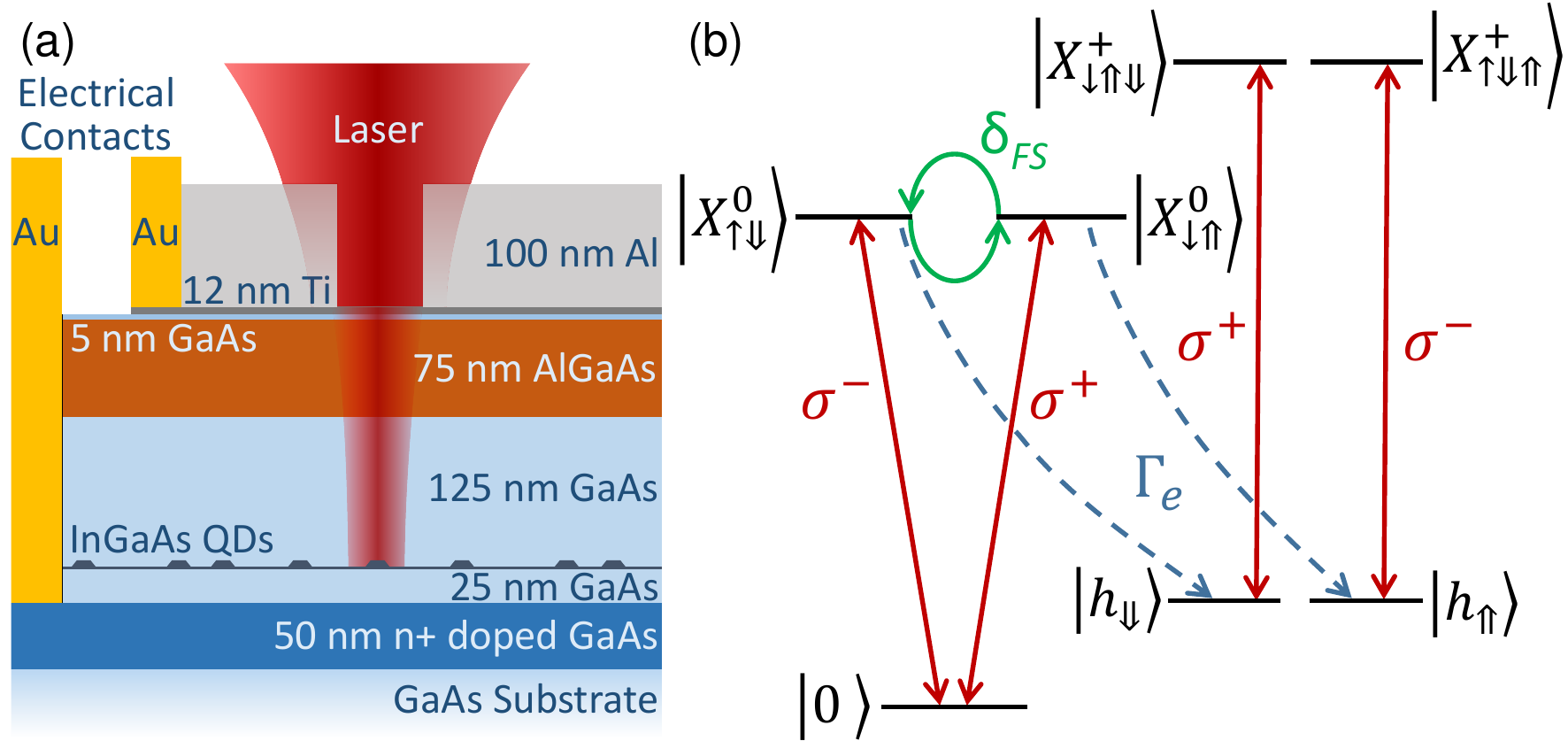}

\protect\caption{\label{fig:1Levels}(a) Sample structure. A low density layer of InGaAs
QDs is embedded in an n-i-Schottky diode. (b) Energy levels in the
circularly polarized basis at zero magnetic field where $(\downarrow/\uparrow)$
and $(\Downarrow/\Uparrow)$ represent electron and hole spins respectively.
The neutral exciton ($X^{0}$) states are coupled by the FSS with
angular precession frequency $\delta_{FS}$ (green arrows) and decay
by electron tunneling at a rate $\Gamma_{e}$ (blue dashed arrows)
to leave single holes ($h$). The hole spin state can be read out
by probing the $h\rightarrow X^{+}$ (positive trion) transitions
using $\sigma^{+/-}$ polarized pulses.}
\end{figure}

When driven by ultrafast pulsed lasers, exciton ionization schemes
can offer both picosecond initialization times and on-demand operation.
Unfortunately, the anisotropic exchange interaction \citep{Gammon1996,Bayer2002}
typically reduces fidelity by causing spin precession during the exciton
lifetime \citep{PhysRevLett.100.197401,:/content/aip/journal/apl/97/6/10.1063/1.3476353,PhysRevB.85.155310}
{[}see Fig. \ref{fig:1Levels}(b){]}. Fast electron tunneling minimizes
this effect with fidelities of $F>96\%$ obtained for ionization in
QD molecules \citep{PhysRevB.85.241306} and $F>97\%$ for probabilistic
(continuous-wave (CW)) initialization of single QDs \citep{PhysRevB.90.241303}.
However, a negative consequence is the reduction of the hole qubit's
lifetime to 300 ps \citep{PhysRevB.85.241306} or 3 ns \citep{PhysRevB.90.241303}
respectively. This is significantly less than the hole's long extrinsic
coherence time ($T_{2}^{*}\simeq$10 ns) \citep{Brunner03072009,DeGreve2011,Greilich2011,PhysRevLett.108.017402},
reducing the coherence time ($T_{2}$) and the number of possible
gate operations. Application of a strong out-of-plane magnetic field
inhibits spin precession resulting in $F>99\%$ \citep{:/content/aip/journal/apl/97/6/10.1063/1.3476353};
however out of plane fields are incompatible with present coherent
control schemes \citep{Brunner03072009,DeGreve2011,PhysRevLett.108.017402,Hansom2014}
which require in-plane spin quantization.

In this Rapid Communication we demonstrate $F>99\%$ at zero magnetic
field with on-demand, < 100 ps initialization and a hole lifetime
that can be as high as 25.2 ns. This is achieved by exciton ionization
in a QD with near-zero fine-structure splitting (FSS), rendering the
precession due to the anisotropic exchange interaction negligible
relative to the exciton lifetime. To demonstrate that such a scheme
is also applicable to typical QDs with finite FSS, we use the optical
Stark effect (OSE) \citep{PhysRevLett.92.157401} to reduce the FSS
\citep{PhysRevLett.100.177401,PhysRevLett.103.217402}, resulting
in increased fidelity.

The sample consists of InGaAs/GaAs self-assembled QDs embedded in
the intrinsic region of an n-i-Schottky diode {[}see Fig. \ref{fig:1Levels}(a){]}.
Five QDs with FSS ranging from 2.01 $\upmu$eV (QD A) to 31.2 $\upmu$eV
(QD E) were studied. The sample is held in a helium bath cryostat
at 4.2 K and excited by transform-limited FWHM $\simeq0.2$ meV pulses
derived from a Ti:Sapphire laser with 76 MHz repetition rate. Photoexcited
carriers in the QD are then detected by measuring the resulting photocurrent
\citep{Zrenner2002}.

Figure \ref{fig:1Levels}(b) illustrates the principle of the hole
spin initialization scheme. A circularly-polarized laser pulse with
$\pi$ pulse area creates a neutral exciton ($X^{0}$) in the QD at
time $t=0$. Under a reverse bias DC electric field ($E$) the exciton
population decays at a rate $\Gamma_{X}=\Gamma_{r}+\Gamma_{e}+\Gamma_{h}$
where $\Gamma_{r}$ is the rate of radiative recombination and $\Gamma_{e}$
and $\Gamma_{h}$ are the electron and hole tunneling rates respectively.Owing
to the larger hole effective mass the electron tunneling rates exceed
hole tunneling rates by around two orders of magnitude ($\Gamma_{e}\gg\Gamma_{h}$).
Radiative recombination rates are slow compared to electron tunneling
in our devices \citep{PhysRevB.70.033301,PhysRevB.77.245311} and
hence $\Gamma_{X}\simeq\Gamma_{e}$. The tunneling of the electron
leaves behind a single hole with spin conserved from the $X^{0}$;
thus the initialization time for the hole is equal to $1/\Gamma_{e}$.
The anisotropic exchange interaction causes precession between $X_{\uparrow\Downarrow}^{0}$
and $X_{\downarrow\Uparrow}^{0}$ states at angular frequency $\delta_{FS}$,
reducing the polarization of the resultant hole spin.

\begin{figure}
\includegraphics[width=1\columnwidth]{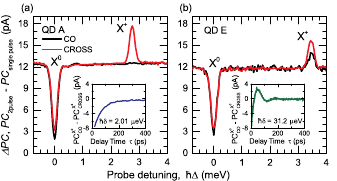}

\protect\caption{\label{fig:PCSpectra}Two-color pump-probe photocurrent spectra of
quantum dots exhibiting (a) negligible (2.01 $\upmu$eV) and (b) large
(31.2 $\upmu$eV) FSS. Spectra are measured at $E=72$ kV $\mathrm{cm}{}^{-1}$
and $\tau=100\:\mathrm{ps}$ when only the hole is left in the QD.
Black (red) lines correspond to a co (cross)-polarized probe laser.
Insets: The precession of the neutral exciton spin measured by time-resolved
pump-probe spectroscopy \citep{1742-6596-245-1-012010}. The exponential
damping of the fine-structure beats corresponds to the exciton lifetime
$(1/\Gamma_{X})$.}
\end{figure}

To measure the initialized hole spin, a co (cross)-circularly polarized
probe $\pi$ pulse arrives after a delay ($\tau$) with a detuning
of $\Delta$ relative to the first pulse. By scanning the probe detuning,
two-pulse spectra like those shown in Fig. \ref{fig:PCSpectra} are
obtained where black (red) traces represent the co (cross)-polarized
cases respectively . For presentation purposes, a single-pulse (probe
only) spectrum is subtracted from the two-pulse spectrum to remove
any weak spectral features not arising from the pumped QD; the dip
at $\Delta=0$ corresponds to subtraction of the $X^{0}$ peak. At
$\Delta$ equal to the positive trion ($X^{+}$) binding energy the
$h\rightarrow X^{+}$ transitions shown in Fig. \ref{fig:1Levels}(b)
are probed. Peaks corresponding to these transitions are observed
in the spectra and the hole spin state may be extracted from their
relative amplitudes.

Figure \ref{fig:PCSpectra}(a) shows the spectrum of QD A with a small
FSS of 2.01 $\upmu$eV. The inset illustrates that exciton spin precession
during electron tunneling is negligible; as a result, the hole spin
preparation is almost ideal with no trion peak observed for a co-polarized
probe. By contrast, Fig. \ref{fig:PCSpectra}(b) shows the case of
QD E with a large FSS of 31.2 $\upmu$eV. The exciton spin precession
is seen clearly in the inset whilst prominent trion peaks in both
spectra illustrate the reduced fidelity.

The fidelity \citep{1994JMOp...41.2315J} of spin preparation is defined
as $F=\left\langle \Uparrow\left|\rho\right|\Uparrow\right\rangle $
where $\rho$ is the density matrix of the prepared spin state and
$\Uparrow\:(\Downarrow)$ is the target spin state. Fidelity is evaluated
using Eq. \ref{eq:PCFidelity}:

\begin{equation}
F=\frac{PC_{\mathrm{cross}}^{X^{+}}}{PC_{\mathrm{cross}}^{X^{+}}+PC_{\mathrm{co}}^{X^{+}}},\label{eq:PCFidelity}
\end{equation}
where $PC_{cross}^{X^{+}}$ and $PC_{co}^{X^{+}}$ are the amplitudes
of the $X^{+}$ peaks in the co- and cross-polarized spectra.

\begin{figure}
\includegraphics[width=1\columnwidth]{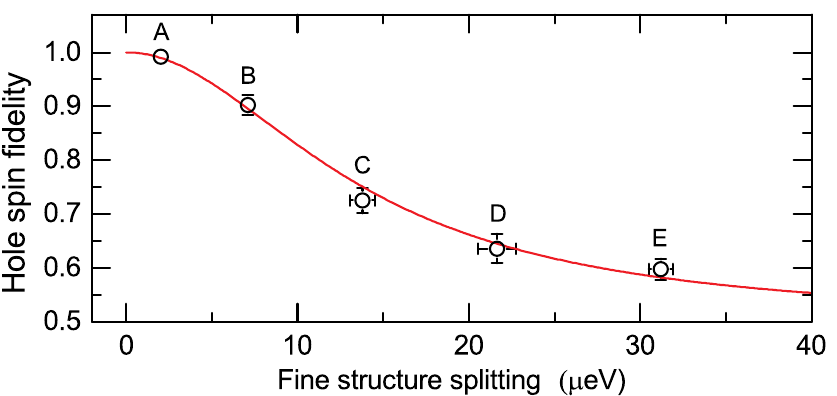}

\protect\caption{\label{fig:FidDependence}Fidelity vs. fine-structure splitting as
measured for QDs with different FSS. For all QDs, \emph{F} was measured
at $\Gamma_{e}=0.021\:\mathrm{ps^{-1}}$. The red line is a calculation
using Eq. \ref{eq:Fidelity}.}
\end{figure}

To investigate the variation of $F$ with $\delta_{FS}$, the fidelities
of the five QDs with different FSS are measured at a constant electron
tunneling rate $\left(\Gamma_{e}\right)$ by varying the DC electric
field. The tunneling rates and FSS are measured by time-resolved pump-probe
spectroscopy \citep{PhysRevB.75.193306,1742-6596-245-1-012010} whilst
the smallest FSS are measured with a narrow linewidth (FWHM $<$ 10
neV) CW laser {[}see Ref. \onlinecite{SMnote}{]}\nocite{wreo2190,Quilter2014, Fry2000,:/content/aip/journal/apl/85/18/10.1063/1.1815373,PSSC:PSSC200671572,PhysRevLett.108.107401,PhysRevB.81.045311,1742-6596-245-1-012010,PhysRevLett.100.177401,Xu2007,PhysRevB.75.193306}.
The data is shown in Fig. \ref{fig:FidDependence} where $F$ falls
as the fine-structure precession increases relative to electron tunneling.
At $B=0\:\mathrm{T}$, the hole spin fidelity is described by a model
developed by Godden et al. \citep{:/content/aip/journal/apl/97/6/10.1063/1.3476353}:

\begin{equation}
F=1-\frac{1}{2}\left[\frac{\delta_{FS}^{2}}{\delta_{FS}^{2}+\left(\Gamma_{X}-\Gamma_{h}\right)^{2}}\right],\label{eq:Fidelity}
\end{equation}
where $\left(\Gamma_{X}-\Gamma_{h}\right)\simeq\Gamma_{e}$ due to
slow radiative recombination as previously discussed.. The line in
Fig. \ref{fig:FidDependence} shows a calculation of $F$ using Eq.
\ref{eq:Fidelity}, demonstrating a good quantitative agreement with
our results. For QD A ($\hbar\delta_{FS}=2.01\pm0.20\:\ensuremath{\upmu}\mathrm{eV}$)
a fidelity lower bound of $F\geq0.993$ is measured. This value is
only limited by the noise present in the co-polarized spectrum and
implies an initialization error rate below the 0.75\% threshold required
for error correction \citep{PhysRevLett.98.190504}.

\begin{figure}
\includegraphics[width=1\columnwidth]{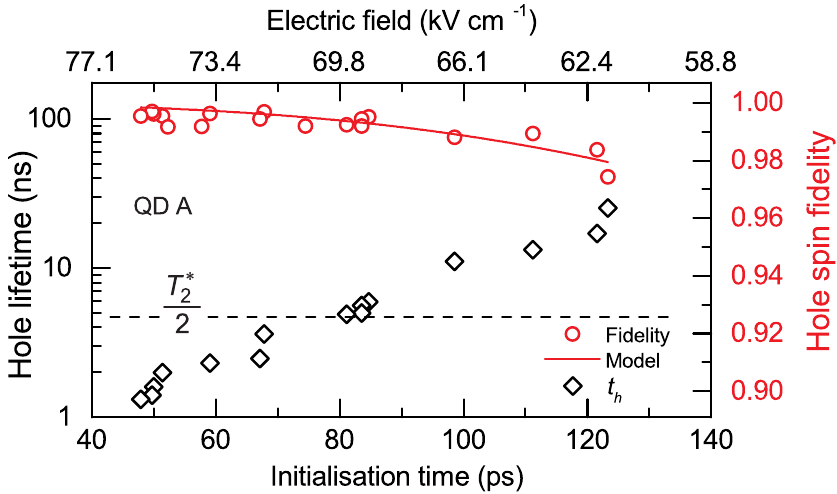}

\protect\caption{\label{fig:FvsLifetime}Hole lifetime ($1/\Gamma_{h}$) (diamonds)
and fidelity lower bound (red circles) plotted vs. initialization
time ($1/\Gamma_{e}$) and approximate DC electric field for QD A.
Error bars are of the order of the data point size. The model (red
line) represents a fit of Eq. \ref{eq:Fidelity} with measured $\hbar\delta_{FS}=2.01\pm0.20\:\ensuremath{\upmu}\mathrm{eV}$
and fitting parameter $\hbar\chi_{E}=-0.0219\pm0.0007\:\ensuremath{\upmu}\mathrm{eV}\:\mathrm{V}^{-1}\:\mathrm{cm}$
corresponding to a small change in FSS with DC electric field {[}see
Ref. \onlinecite{SMnote}{]}.}
\end{figure}

Owing to the negligible exciton spin precession of QD A, fast electron
tunneling is no longer required to achieve high fidelities. This enables
the reduction of the diode electric field to maximize the hole lifetime
($T_{h}=1/\Gamma_{h}$) with a moderate increase in initialization
time ($1/\Gamma_{e}$ which remains $\ll$ than the coherence time
($T_{2}$)) and a small change in FSS \citep{PhysRevB.81.045311}.
Previous studies on similar samples have shown that the coherence
time of the hole spin is limited by the hole tunneling rate \citep{PhysRevLett.108.017402}
($\Gamma_{h}$) at typical electric fields. Beyond this, the next
limit is the extrinsic pure dephasing time ($T_{2}^{\ast}\simeq10\:\mathrm{ns}$
\citep{Brunner03072009,DeGreve2011,Greilich2011,PhysRevLett.108.017402})
which most likely originates from fluctuations in the electric field
acting on the hole g-factor \citep{DeGreve2011,Greilich2011,Kuhlmann2013}.
In the limit of negligible extrinsic pure dephasing, or spin-flips,
the coherence time is twice the hole lifetime. Thus, a good target
is $2T_{h}>T_{2}^{\ast}$, the point at which pure dephasing rather
than hole tunneling becomes the dominant limitation on the coherence
time.

To demonstrate this, Fig. \ref{fig:FvsLifetime} shows the results
of measuring fidelity, initialization time and hole lifetime for a
range of DC electric fields on the low FSS QD A. By treating the variation
of $\delta_{FS}$ with DC electric field as a fitting parameter in
Eq. \ref{eq:Fidelity} {[}see Ref. \onlinecite{SMnote}{]} we obtain
excellent agreement with the data {[}see red line{]}. At lower electric
fields the maximum resolvable fidelity decreases due to reduced photocurrent,
emphasizing that our measurements represent a lower-bound. For $2T_{h}>T_{2}^{\ast}$
the initialization time ranges from 83.5 ps to 123 ps with fidelity
lower bounds from $\geq0.974$ to $\geq0.995$, indicating that both
high fidelity and long qubit lifetimes may be obtained for a QD with
negligible FSS.

Due to the importance of low FSS QDs for polarization entangled photon
sources \citep{Stevenson2006,Akopian2006}, deterministic growth of
symmetric QDs is a topical area of research \citep{:/content/aip/journal/apl/101/2/10.1063/1.4733664,Juska2013,:/content/aip/journal/apl/102/15/10.1063/1.4802088}
but is yet to be demonstrated. As such, in-situ methods for tuning
FSS are widely studied, using strain \citep{:/content/aip/journal/apl/88/20/10.1063/1.2204843,PhysRevLett.109.147401},
magnetic \citep{PhysRevB.73.033306} and laser \citep{PhysRevLett.100.177401,PhysRevLett.103.217402}
fields as well as both lateral \citep{:/content/aip/journal/apl/91/18/10.1063/1.2805025,:/content/aip/journal/apl/90/4/10.1063/1.2431758}
and vertical \citep{Bennett2010,Ghali2012} DC electric fields. In
order to retain control over the qubit energy and lifetime it is desirable
to tune the FSS using a field that is independent from the DC electric
field. Thus we use a detuned CW laser to tune $\delta_{FS}$ by the
OSE \citep{PhysRevLett.92.157401,PhysRevLett.100.177401,PhysRevLett.103.217402}
at a fixed DC electric field.

\begin{figure*}
\includegraphics[width=1\textwidth]{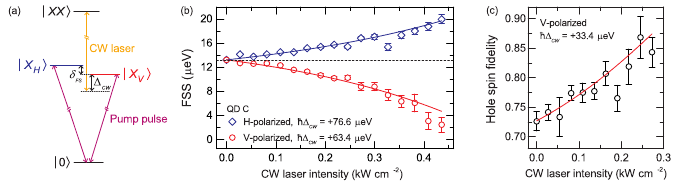}

\protect\caption{(a) QD energy levels in the linear basis where the neutral exciton
eigenstates are split by $\delta_{FS}$. The CW laser is $V$-polarized
and positively detuned from the $X_{V}\rightarrow XX$ transition
by $\triangle_{CW}$. The pump pulse addresses both exciton levels
owing to its circular polarization and FWHM $\gg$ $\delta_{FS}$.
(b) FSS vs. CW laser intensity for QD C with $\hbar\delta_{FS}=13.2\:\ensuremath{\upmu}\mathrm{eV}$
measured at $E=60$ kV $\mathrm{cm}^{-1}$ to resolve small $\delta_{FS}$.
The blue diamonds (red circles) correspond to an $H$ $(V)$-polarized
CW laser which addresses the high (low) energy exciton eigenstate.
The solid lines are fits to Eq. \ref{eq:StarkShift}. (c) Hole spin
fidelity (measured as in Fig. \ref{fig:PCSpectra}) vs. CW laser intensity
for a $V$-polarized CW laser at $E=72$ kV $\mathrm{cm}^{-1}$ to
maximize photocurrent detection efficiency. The line is a fit of the
model (Eq. \ref{eq:Fidelity}) incorporating the variation of $\delta_{FS}$
with CW laser intensity {[}see Ref. \onlinecite{SMnote}{]} .\label{fig:StarkFidelity}}
\end{figure*}

In our scheme the OSE is induced by a linearly polarized CW laser
which is positively detuned from the co-polarized $X\rightarrow XX$
(biexciton) transition by $\Delta_{CW}$ as illustrated in Fig. \ref{fig:StarkFidelity}(a).
The neutral exciton eigenstates ($X_{H/V}$) are linearly polarized
along the in-plane crystal axes and can be addressed individually
by selecting the laser polarization. $X_{H}$ and $X_{V}$ are split
by $\hbar\delta_{FS}$; we define $X_{V}$ to be lower in energy.
A positive-detuned $V$-polarized laser addresses the $X_{V}$ state
and acts to reduce $\delta_{FS}$ by Stark-shifting the $X_{V}$ state
to higher energy. By contrast, an $H$-polarized laser increases $\delta_{FS}$
by shifting the $X_{H}$ state to higher energy. In the case of positive
detuning ($\Delta_{CW}>0$), the change in FSS due to the OSE ($\varDelta\omega$)
is given by \citep{PhysRevLett.92.157401}:

\begin{equation}
\varDelta\omega=\frac{s}{2}\left(\Delta_{CW}-\sqrt{\Delta_{CW}^{2}+\left|\Omega\right|^{2}}\right)\label{eq:StarkShift}
\end{equation}
where $\Omega$ is the Rabi splitting induced by the CW laser (proportional
to the square-root of laser intensity $\sqrt{I}$) and $s=\pm1$ when
the CW laser is $H$/$V$ polarized. Similar to previous reports \citep{PhysRevLett.100.177401,Xu2007},
we observe a polarization-independent blue-shift of the exciton energy
with $I$. This arises due to charge screening from the large number
of carriers generated in the surrounding material by the CW laser
and results in a linear dependence of $\Delta_{CW}$ on $I$.

In Fig. \ref{fig:StarkFidelity}(b) the CW laser photon energy is
fixed and the FSS of QD C is measured by time-resolved pump-probe
spectroscopy {[}as in insets to Fig. \ref{fig:PCSpectra}{]} as a
function of both laser intensity ($I$) and polarization. For a $V$-polarized
CW laser (red circles), $\delta_{FS}$ reduces from its initial value
of $\hbar\delta_{FS}=13.2\pm0.1\:\ensuremath{\upmu}\mathrm{eV}$ to
a minimum of $\hbar\delta_{FS}=2.49\pm1.25\:\ensuremath{\upmu}\mathrm{eV}$
at $I=0.44\:\mathrm{kW\:cm^{-2}}$. Conversely, when the laser is
$H$-polarized (blue diamonds) $\delta_{FS}$ increases, proving that
the change in FSS is induced by the OSE. The solid lines show a fit
of Eq. \ref{eq:StarkShift} to the data {[}see Ref. \onlinecite{SMnote}{]}.

To demonstrate that reducing FSS leads to an increase in fidelity,
hole spin fidelity was measured as a function of CW laser intensity.
The laser is $V$-polarized and $\hbar\triangle{}_{CW}=33.4\:\mathrm{\ensuremath{\upmu}eV}$.
The result of this measurement is shown in Fig. \ref{fig:StarkFidelity}(c);
for $I=0.25\:\mathrm{kW\:cm^{-2}}$ (FSS $\simeq8.7\:\ensuremath{\upmu}\mathrm{eV}$)
a fidelity of $F=0.868\pm0.036$ is measured, an increase of 0.142
over that measured with no CW laser. The red line shows a calculation
using Eq. \ref{eq:Fidelity} with experimentally derived parameters
{[}see details in Ref. \onlinecite{SMnote}{]} which again agrees
closely with the data. In both experiments, the maximum $I$ is limited
by photocurrent fluctuations due to laser power instability. This
particularly limits the fidelity measurement as at $I>0.25\:\mathrm{kW\:cm^{-2}}$
fluctuations exceed the small ($\sim1\:\mathrm{pA}$) co-polarized
peak amplitude, limiting the maximum $F$ that can be measured. However,
the agreement with the model and the large optical Stark shift observed
in Fig. \ref{fig:StarkFidelity}(b) indicate that fidelities as high
as those measured for QD A could in principle be obtained with this
method. We also note that the anti-crossing behavior seen with tuning
methods such as strain \citep{PhysRevB.83.121302} does not occur
for the OSE \citep{PhysRevLett.100.177401}.

In conclusion, we have demonstrated that a QD with very small FSS
($2.01\pm0.20\:\ensuremath{\upmu}\mathrm{eV}$) enables fast, on-demand
initialization of a long-lived ($2T_{h}>T_{2}^{\ast}\sim10\:\mathrm{ns}$)
hole spin qubit with fidelity $\geq$99.5\% at $B=0$ T, exceeding
the threshold required for a fault-tolerant QIP implementation \citep{PhysRevLett.98.190504}.
Whilst the high fidelities here are measured at zero magnetic field,
we note that simulations with small $\delta_{FS}$ show that $F$
will remain very high even under the presence of a modest in-plane
magnetic field \citep{PhysRevB.85.155310}. As a result, this initialization
scheme offers performance compatible with coherent control of hole
spins \citep{Brunner03072009,DeGreve2011,PhysRevLett.108.017402,Hansom2014}
where fast gate times ($\sim$20 ps \citep{DeGreve2011,PhysRevLett.108.017402}),
high gate fidelities (94.5\% \citep{DeGreve2011}) and long coherence
lifetimes have demonstrated an attractive qubit platform. We note
that hole lifetimes could be further extended by modulation of the
electric field to a very low value between initialization and readout
\citep{:/content/aip/journal/apl/102/18/10.1063/1.4804373}. Combining
this with supression of extrinsic pure dephasing by optical spin echo
\citep{Press2010} could enable high fidelity initialization with
coherence times in the $\ensuremath{\upmu}\mathrm{s}$ regime.

Furthermore, we have demonstrated that the initialization fidelity
for arbitrary QDs with larger FSS can be increased by the OSE, providing
additional motivation for FSS tuning studies \citep{PhysRevLett.109.147401,PhysRevB.73.033306,PhysRevLett.100.177401,PhysRevLett.103.217402,:/content/aip/journal/apl/91/18/10.1063/1.2805025,:/content/aip/journal/apl/88/20/10.1063/1.2204843,Bennett2010,Ghali2012,:/content/aip/journal/apl/90/4/10.1063/1.2431758}
that were typically motivated by the generation of entangled photon
pairs \citep{Stevenson2006,Akopian2006}. In our devices the DC electric
field presents an extra tunable parameter that may be used to optimize
qubit lifetimes {[}see Fig. \ref{fig:FvsLifetime}{]} or to tune two
QDs into resonance. This presents a potential route towards fault-tolerant
QIP schemes based on multiple long-lived hole spins on a single chip.

\emph{Note added in proof.} Recently, we became aware of related results
by another group \citep{PhysRevB.92.115306}.
\begin{acknowledgments}
This work was funded by the EPSRC (UK) programme grant EP/J007544/1.
The authors thank H. Y. Liu and M. Hopkinson for sample growth and
P. Kok for helpful discussions.
\end{acknowledgments}

\appendix
\clearpage
\onecolumngrid
\begin{center} 
\textbf{\large Supplemental Materials: High-fidelity initialization of long-lived quantum dot hole spin qubits by reduced fine-structure splitting} 
\end{center} 
\setcounter{equation}{0} 
\setcounter{figure}{0} 
\setcounter{table}{0} 
\setcounter{page}{1} 
\makeatletter 
\renewcommand{\theequation}{S\arabic{equation}}
\renewcommand{\thefigure}{S\arabic{figure}} 
\renewcommand{\thetable}{S\arabic{table}}


\subsection{Measurements of fidelity}

Fidelity is measured according to Eq. 2 in the body of the paper.
For very high fidelities, it is no longer reliable or meaningful to
fit a Gaussian peak to the co-polarized spectrum. In this case we
use the variance of the photocurrent noise to estimate the amplitude
\citep{wreo2190} and report a lower bound. We sample the datapoints
within the laser FWHM of the trion energy and then calculate the amplitude
estimate ($\epsilon$) as in Eq. \ref{eq:HighFEst}:
\begin{equation}
\epsilon=\frac{\sigma}{\sqrt{N}},\label{eq:HighFEst}
\end{equation}
where $N$ is the number of datapoints within the sample and $\sigma$
is their standard deviation. The quantity $\epsilon$ then replaces
$PC_{co}^{X^{+}}$ in Eq. 2 to calculate the lower bound of $F$.

\subsection{Conversion of biased voltage to DC electric field\label{sub:Conversion-of-biased}}

The application of a reversed biased voltage ($V$) generates a direct
current (DC) electric field ($E$) inside the diode structure. The
voltage can be converted to electric field accroding to \citep{Quilter2014}:
\begin{equation}
E=(V+V_{bi})/W_{i},\label{eq:VtoE}
\end{equation}
where $V_{bi}$ ($\sim0.76$ V) is the built-in voltage of the diode.
$W_{i}$ (230 nm) is the distance between the Ohmic and Schottky contacts.

\subsection{Measurement of small FSS}

\begin{figure}[H]
\begin{centering}
\includegraphics[width=0.5\columnwidth]{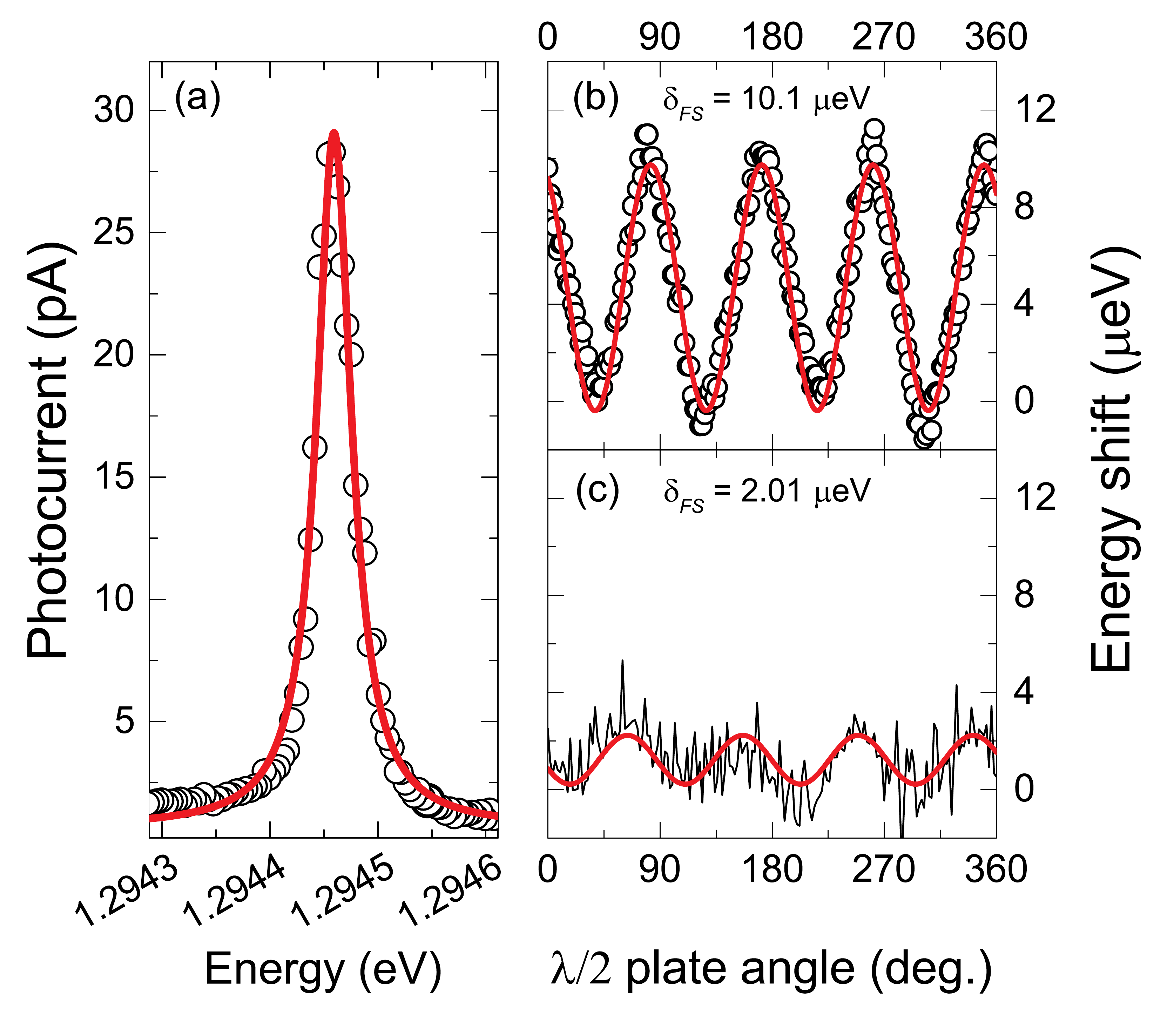}
\par\end{centering}

\protect\caption{\label{fig:CWFSS}(a) Typical high-resolution photocurrent spectrum
of a QD with Lorentzian fit (red line) of linewidth $\mathrm{FWHM}=39.3\pm0.3\:\ensuremath{\upmu}\mathrm{eV}$.
Plot of the neutral exciton energy of QDs with (b) moderate $\delta_{FS}$
and (c) very small $\delta_{FS}$ (QD A) as a function of half-wave
plate angle. Red lines: Fitting with $\sin^{2}\left(\theta\right)$
function. The amplitude of the fits yields a fine structure splitting
of $\hbar\delta_{FS}=10.1\pm0.1\:\ensuremath{\upmu}\mathrm{eV}$ and
$2.01\pm0.2\:\ensuremath{\upmu}\mathrm{eV}$ respectively.}
\end{figure}

In order to measure small FSS we use a narrow linewidth CW laser (FWHM
$\sim$ 1 MHz) operating at a fixed wavelength to perform high resolution
photocurrent spectroscopy \citep{Fry2000,:/content/aip/journal/apl/85/18/10.1063/1.1815373,PSSC:PSSC200671572}.
The $0\rightarrow X$ transition is Stark-shifted through the laser
line by changing the applied DC electric field. After conversion from
DC electric field to energy, the result is a Lorentzian lineshape
with typical linewidth $\sim40\:\ensuremath{\upmu}\mathrm{eV}$ as
illustrated in Fig. \ref{fig:CWFSS}. The linewidth broadening relative
to the exciton decay time limit of ($\sim14\:\ensuremath{\upmu}\mathrm{eV}$
determined by the electron tunnelling rate) is attributed to charge
fluctuations in the electrostatic environment of the dot during the
measurement \citep{PhysRevLett.108.107401}.

Rotating the linear polarization angle of the CW laser with a half-wave
plate causes the exciton energy to oscillate with an amplitude of
$\hbar\delta_{FS}$ as illustrated for a QD with moderate $\delta_{FS}$
$(10.1\pm0.1\:\ensuremath{\upmu}\mathrm{eV)}$ and very small $\delta_{FS}$
($2.01\pm0.20\:\ensuremath{\upmu}\mathrm{eV}$) (QD A) in Fig. \ref{fig:CWFSS}(b)
and (c).

\subsection{Model of fidelity vs. applied DC electric field}

\begin{figure}
\includegraphics[width=0.45\columnwidth]{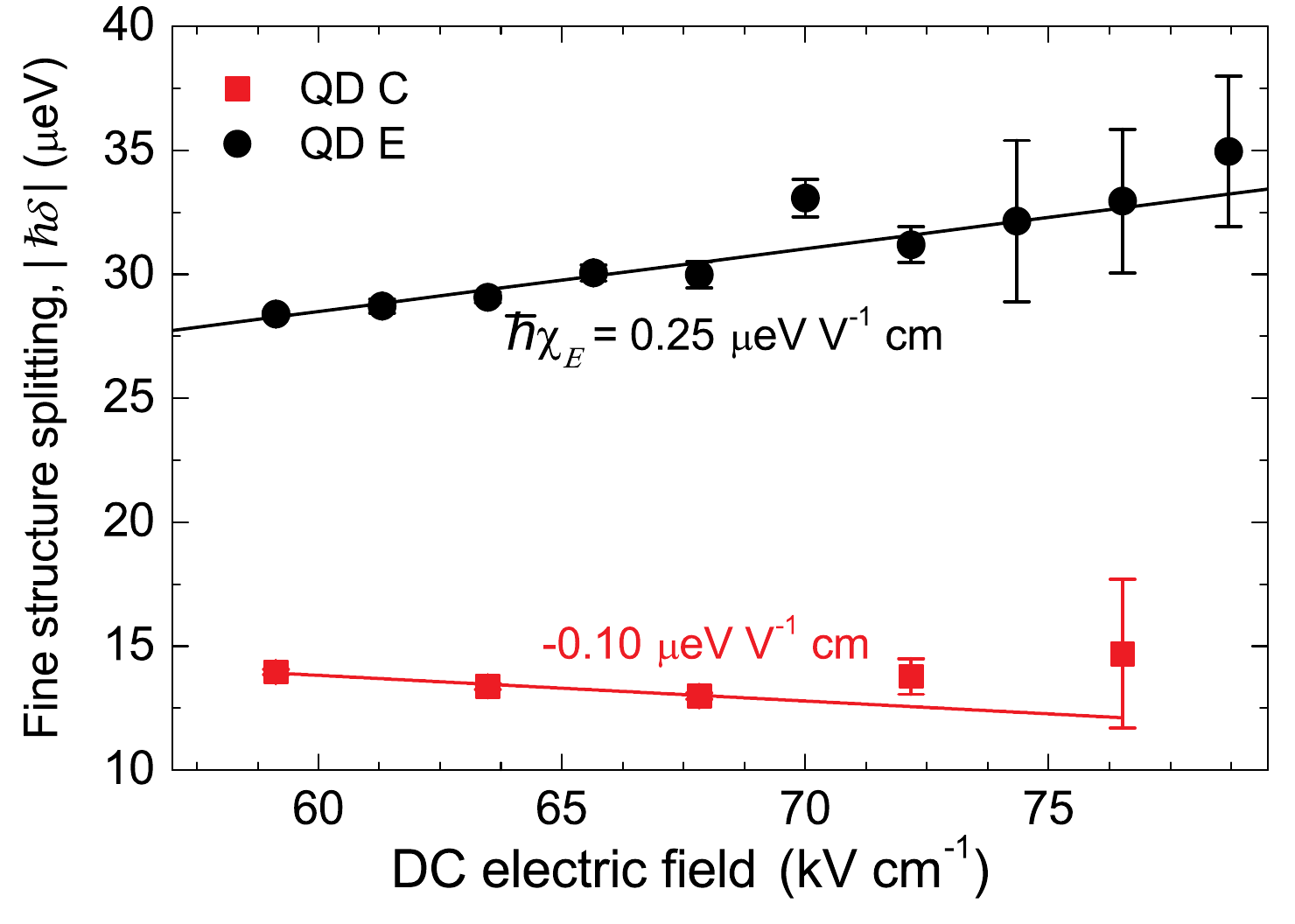}

\protect\caption{Variation of FSS with DC electric field for QDs C and E with different
values of \foreignlanguage{british}{$\delta_{FS}$}. The labels show
the values of $\hbar\chi_{E}$ extracted from the linear fits. The
opposite signs result from the two QDs being elongated along orthogonal
crystal axes whilst the deviation of the red data points from the
fit at high DC electric field corresponds to the onset of $\Gamma_{X}\gg\delta_{FS}$
and thus the resolution limit for a given QD and DC electric field.
This figure is adapted from Ref. \citep{Quilter2014}.\label{fig:FSSvarV}}
\end{figure}

The key modification to the model of Eq. 1 for experiments at non-constant
DC electric field {[}see Fig. 4 in the main paper{]} is to consider
the variation of FSS with DC electric field \citep{PhysRevB.81.045311}.
In this case the expression for FSS is defined by Eq. \ref{eq:FSSV}:

\begin{equation}
\delta_{FS}\left(E\right)=\delta_{FS}\left|_{_{E{}_{0}}}\right.+\chi_{E}\left[E-E_{0}\right],\label{eq:FSSV}
\end{equation}
where $\delta_{FS}\left|_{_{E_{0}}}\right.$ is the FSS evaluated
at $E_{0}$ and $\chi_{E}$ is a linear gradient of FSS with $E$.
The linear gradient represents a good approximation of the form of
$\chi_{E}$ however full calculations require numerical methods \citep{PhysRevB.81.045311}.
For larger FSS we can simply measure $\hbar\chi_{E}$ {[}as shown
in Fig. \ref{fig:FSSvarV}{]} using time-resolved pump-probe measurements
{[}as shown in the insets to Fig. 2 of the main paper{]} . The fittings
give $\hbar\chi_{E}=0.25\pm0.04\:\mathrm{\ensuremath{\upmu}eV\:V^{-1}\:cm}$
for QD E and $-0.10\pm0.02\:\mathrm{\ensuremath{\upmu}eV\:V^{-1}\:cm}$
for QD C. These values are consistent with the literature value (0.285
$\mu$eV $\mathrm{V}{}^{-1}$ cm) reported by Bennett et al.$^{34}$
and align with the theoretical prediction that smaller values of $\delta_{FS}$
(i.e. initial QD anisotropy) give smaller values of $\chi_{E}$ \citep{PhysRevB.81.045311}.

As such, we expect a very small value of $\hbar\chi_{E}$ for QD A
($\hbar\delta_{FS}=2.01\pm0.20\:\ensuremath{\upmu}\mathrm{eV}$).
Since the FSS of QD A is already more than an order of magnitude smaller
than the QD linewidth {[}see Fig. \ref{fig:CWFSS}{]}, $\chi_{E}$
is used as a fitting parameter. Inserting Eq. \ref{eq:FSSV} into
the model of Godden et al. \citep{:/content/aip/journal/apl/97/6/10.1063/1.3476353}
{[}Eq. \ref{eq:Fidelity} in the main text{]} and fitting $\chi_{E}$
produces the line shown in Fig. \ref{fig:FvsLifetime} of the paper.
The extracted value of $\hbar\chi_{E}=-0.0219\pm0.0007\:\ensuremath{\upmu}\mathrm{eV}\:\mathrm{V}^{-1}\:\mathrm{cm}$
is physically reasonable and again agrees with the expected trend
between $\delta_{FS}$ and $\chi_{E}$.

\subsection{Fine structure splitting vs. CW laser intensity}

\begin{figure}
\includegraphics[scale=0.5]{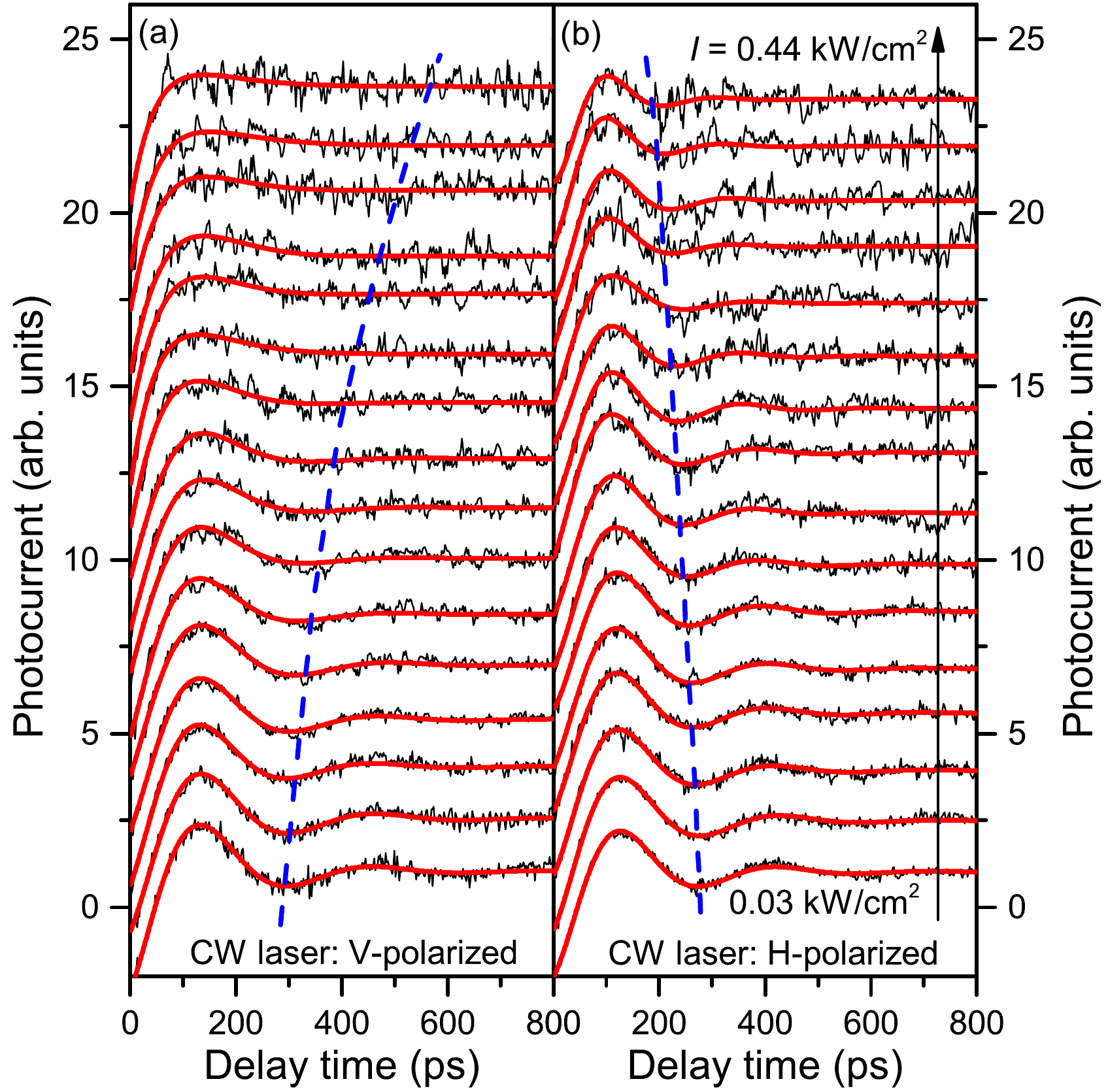}

\protect\caption{Fine structure precession of the exciton spin vs. CW laser intensity
\foreignlanguage{british}{$I$} measured by time-resolved pump-probe
photocurrent technique. The CW laser is either (a) \foreignlanguage{british}{$V$}-
or (b) \foreignlanguage{british}{$H$}-polarized and positively detuned
from the \foreignlanguage{british}{$X\rightarrow XX$} transition
(see Fig. 5(a) in the main paper). Detuning = 76.6 and 63.4 $\ensuremath{\upmu}\mathrm{eV}$
when the CW laser is \foreignlanguage{british}{$H$/$V$-}polarized
respectively. The CW laser intensity ranges from 0.03 to 0.44 kW/$\mathrm{cm{}^{2}}$.
Red lines: Fitting with an exponentially damped sine function. Blue
lines: guides for the eye.\label{fig:FSSvsI}}
\end{figure}

To demonstrate the tuning of $\delta_{FS}$ using OSE, we measured
the FSS of dot C by time-resolved pump-probe \citep{1742-6596-245-1-012010}
with an additional tuneable narrowband CW laser incident on the sample.
The CW laser which is $H$/$V$-polarized and positively deutned from
the $X\rightarrow XX$ transition {[}see Fig. 5(a) in the main paper{]}
is used to tune the FSS. Fig. \ref{fig:FSSvsI} shows the fine structure
precession of the exciton spin vs. the CW laser intensity $I$. Since
the smallest $\delta_{FS}$ that can be resolved by this measurement
is limited by the exciton decay time, a relatively low DC electric
field ($E=60$ kV cm$^{-1}$) was used. When the CW laser is $V$-polarized
{[}see Fig. \ref{fig:FSSvsI}(a){]}, the frequency of the fine structure
precession decreases with the CW laser intensity. At $I$ = 0.44 kW
$\mathrm{cm{}^{-2}}$, no fine structure precession is observed, indicating
that a very small FSS close to the resolution limit of the time-resolved
pump-probe measurement is achieved. By contrast, the frequency of
the fine structure precession increases with CW laser intensity when
the CW laser is $H$-polarized {[}see Fig. \ref{fig:FSSvsI}(b){]},
verifying that the change of the FSS is induced by OSE. $\delta_{FS}$
can be extracted by fitting the data with an exponentionally damped
sine function {[}see red lines{]}. The $\delta_{FS}$ vs. the CW laser
intensity is shown in Fig. 5(b) in the main paper.

\subsection{Model of fidelity vs. OSE}

The increase of the hole spin fidelity by reducing the FSS using OSE
{[}see Fig. \ref{fig:StarkFidelity}(c) in the main paper{]} can be
well described by incorporting the OSE into the model of Godden et
al. \citep{:/content/aip/journal/apl/97/6/10.1063/1.3476353} (see
Eq. \ref{eq:Fidelity} in the main text). The FSS with the presence
of a CW laser postivltiy detuend from the $X\rightarrow XX$ transition
{[}see Fig. \ref{fig:StarkFidelity}(a) in the main paper{]} is given
by:

\begin{equation}
\delta_{FS}(I)=\left.\delta_{FS}\right|_{_{I=0}}+\varDelta\omega,\label{eq:FSSvI}
\end{equation}
\begin{equation}
\varDelta\omega=\frac{s}{2}\left(\Delta_{CW}-\sqrt{\varDelta_{CW}^{2}+\left|\Omega\right|^{2}}\right),\label{eq:FSSDeltaW}
\end{equation}
where $I$ is the CW laser intensity. $\varDelta\omega$ is the change
of the FSS induced by OSE \citep{PhysRevLett.92.157401}. $\varDelta_{CW}$
is the detuning of the CW laser. $s=\pm1$ when the CW laser is $H$/$V$
polarized. $\Omega=\sqrt{aI}/\hbar$ is the Rabi splitting induced
by the CW laser. $a$ is a fitting parameter proportional to the optical
dipole momentum of the $X\rightarrow XX$ transition. In these experiments
a linear blue-shift of the $0\rightarrow X_{H/V}$ transitions with
laser intensity is observed when the CW laser is applied \citep{PhysRevLett.100.177401,Xu2007}.
This effect is independent of laser polarization, we thus attribute
the shift to charge screening from the large number of carriers generated
in the surrounding material by the CW laser as in previous studies
\citep{PhysRevLett.100.177401,Xu2007}. A similar blue shift is expected
for the $X\rightarrow XX$ transition; hence $\Delta_{CW}$ is dependent
on the incident CW laser intensity ($I$) according to:

\begin{equation}
\Delta_{CW}\left(I\right)=\left.\Delta_{CW}\right|_{_{I=0}}-kI/\hbar,\label{eq:DeltaCW}
\end{equation}
where $k$ is a fitting parameter. Fig. \ref{fig:StarkFidelity}(b)
in the main paper shows $\delta_{FS}$ vs. the CW laser intensity
measured at $E=6$0 kV cm$^{-1}$ and the fits according to Eq. \ref{eq:FSSvI}.
The parameters used in these fits are shown in Table \ref{tab:ParamsFSS}.

\begin{table}
\begin{tabular}{|c|c|c|c|c|c|}
\hline 
Polarization & $\hbar\delta_{FS}|_{I=0}$($\ensuremath{\upmu}$eV) & $s$ & $a$ (meV$^{2}$$\ensuremath{\upmu}\mathrm{m}^{2}$ W$^{-1}$) & $\hbar\triangle_{CW}|_{I=0}$ ($\ensuremath{\upmu}$eV) & $k$ (eV $\ensuremath{\upmu}\mathrm{m}^{2}$ W$^{-1}$)\tabularnewline
\hline 
H & 13.2 & $+1$ & 275 & 76.6 & 8.4\tabularnewline
\hline 
V & 13.2 & $-1$ & 275 & 63.4 & 8.4\tabularnewline
\hline 
\end{tabular}\protect\caption{\label{tab:ParamsFSS}Parameters used in the fits of FSS vs. CW laser
power {[}see Fig. 5(b) in the main paper{]}.}
\end{table}

Knowing how the FSS depends on the CW laser intensity, we now discuss
the fidelity of the hole spin initialization vs. the CW laser intensity.
To demonstrate the increase of the hole spin fidelity by reducing
the FSS using OSE, the hole spin fidelity was measured as a function
of the CW laser intensity {[}see Fig. \ref{fig:StarkFidelity}(c)
in the main paper{]}. This data can be well reproduced by including
Eq. \ref{eq:FSSvI} in Eq. \ref{eq:Fidelity} in the main paper:

\begin{equation}
F=1-\frac{1}{2}\left[\frac{(\left.\delta_{FS}\right|_{_{I=0}}+\varDelta\omega)^{2}}{(\left.\delta_{FS}\right|_{_{I=0}}+\varDelta\omega)^{2}+\left(\Gamma_{X}-\Gamma_{h}\right)^{2}}\right],
\end{equation}
where $\Gamma_{X}$ and $\Gamma_{h}$ are the exciton and hole decay
rate. Table \ref{tab:ParamsF} lists the parameters used in this fit.
$a$ is determined from the fit of the $\delta_{FS}$ vs. CW laser
intensity measured at $E=60$ kV cm$^{-1}${[}see Fig. 5(b) in the
main paper{]}. $\Gamma_{X}-\Gamma_{h}$ is determined from time-resolved
pump-probe spectroscopy \citep{PhysRevB.75.193306} as discussed in
the main paper.

\begin{table}
\begin{tabular}{|c|c|c|c|c|c|c|}
\hline 
Polarization & $\hbar\delta_{FS}|_{I=0}$($\ensuremath{\upmu}$eV) & $s$ & $a$ (meV$^{2}$$\ensuremath{\upmu}\mathrm{m}^{2}$ W$^{-1}$) & $\hbar\triangle_{CW}|_{I=0}$ ($\ensuremath{\upmu}$eV) & $k$ (eV $\ensuremath{\upmu}\mathrm{m}^{2}$ W$^{-1}$) & $\Gamma_{X}-\Gamma_{h}$(ps$^{-1}$)\tabularnewline
\hline 
V & 13.2 & $-1$ & 275 & 33.4 & 3.5 & 0.021\tabularnewline
\hline 
\end{tabular}\protect\caption{\label{tab:ParamsF}Parameters used in the fit of hole spin fidelity
vs. CW laser power {[}see Fig. 5(c) in the main paper{]}.}
\end{table}

\end{document}